\documentclass{article}
\usepackage[T1]{fontenc} 
\usepackage[utf8]{inputenc} 
\usepackage{amsmath,cite,url}
\usepackage{graphicx}
\usepackage{color}
\usepackage{enumitem}
\usepackage{lineno}

\title{A Computational Analysis of Pitch Drift in Unaccompanied Solo Singing using DBSCAN Clustering}

\author
{S. Shafiei and S. Hakam\\
  \texttt{sepideh.shafiee@gmail.com}}

\begin{document}

\maketitle
\begin{abstract}
\par Unaccompanied vocalists usually change the tuning unintentionally and end up with a higher or lower pitch than the starting point during a long performance. This phenomenon is called pitch drift, which is dependent on various elements, such as the skill of the performer, and the length and difficulty of the performance. In this paper, we propose a computational method for measuring pitch drift in the course of an unaccompanied vocal performance, using pitch histogram and DBSCAN clustering.\footnote{ The software is available at https://github.com/SepiSha/PitchDrift} 
\end{abstract}

\section{Introduction}

Intonation drift is an important topic in analyzing unaccompanied singing and has been discussed in MIR literature from various angles. In order to study the drift, we need to assume a reference pitch, and then measure the deviation from that reference pitch in the course of a segment or performance. Intonation drift happens both in choirs and solo singing, and it usually occurs in the downward direction [1], [2], and [3]. Harmonic progression has been mentioned as one of the causes of drift in choir [4] and [5]. In this paper, we consider drift in solo singing. Our main goal is to computationally measure the drift in the course of a performance. 
\par
The methodology is discussed in sections 2 to 5. In order to explain the detail we have used an example of a performance \footnote{ The mp3 files for the performance in Example 1 is available in the github repository of the software along with some other examples from various musical cultures.}  with duration 5':32'' (Example 1). It starts with an introduction for about 48 seconds, before the main part starts at the second 00:52. The whole piece has sixteen sentences, which are separated with long silences. The first two sentences are the introduction. A transcription of the whole performance can be found in the corresponding author's dissertation ([6]: 69-71).

\section{Pitch Recognition}
\par There are different algorithms for pitch recognition. All these algorithms use the fundamental frequency of the sound to quantify pitch. Fundamental frequency is the lowest frequency of the sound wave and corresponds to its most dominant perceived pitch. Monophonic voice pitch estimation algorithms have been evaluated and discussed extensively in MIR literature. Gomez et al. have evaluated both algorithms CREPE [7] and pYIN [8] as state-of-the-art pitch recognition methods for monophonic voice [9]. They have evaluated the results of both of these algorithms on iKala dataset and obtained almost similar results for both pYIN and CREPE: 91\% of Raw Pitch Accuracy for pYIN for monophonic voice and 90.5\% accuracy for CREPE. 
We used pYIN with Sonic Annotator for pitch recognition with the following parameters: step size of 256, block size of 2048, low amplitude suppression of 0.1, onset sensitivity of 0.7, prune threshold of 0.1, and threshold distribution of 2. Figure 1 shows the the time-frequency graph of Example 1.

\begin{figure}
 \centerline{
 \includegraphics[width=1\columnwidth]{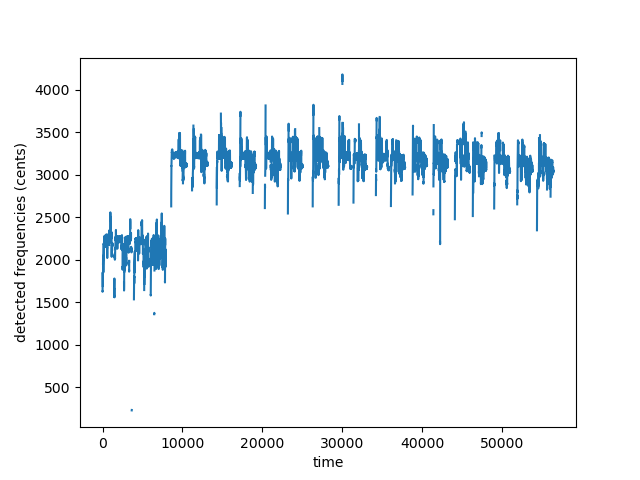}}
 \caption{time-frequency graph for Example 1}
 \label{fig:example}
\end{figure}

\section{Identifying Musical Phrases}
\par Since we are dealing with voice, we do the segmentation based on the relatively long silences between the sentences performed by the vocalist. This method helps us to identify musical phrases (sentences).  An alternative for the purpose of computing pitch drift is to consider congruent segments, i.e. every $n$ seconds. Other attributes that one might consider are intonation drift per $n$ seconds, intonation drift per note, and intonation drift per musical phrases.

\section {Pitch Histogram and its Peaks}

The next step is to find the histograms of the performed pitches for each segment. Pitch histogram has been used before by Bozkurt for analysis of Turkish Makam Music [10]. The use pitch histogram in finding intervals of a performed vocal piece has been discussed extensively in [11] and [12].  Figure 2 shows the pitch histogram of Example 1. 

\begin{figure}
 \centerline{
 \includegraphics[width=1\columnwidth]{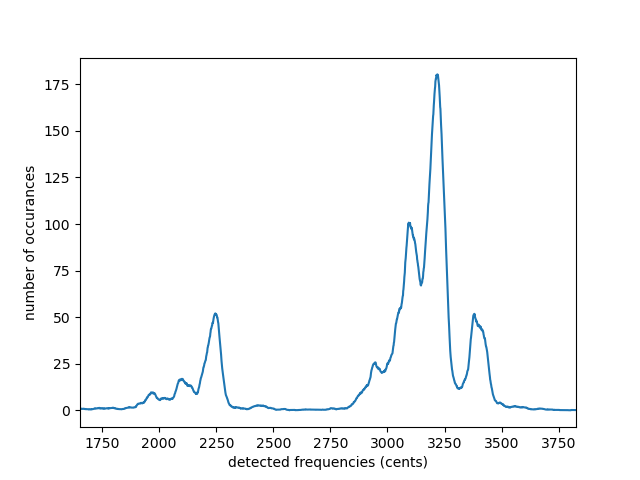}}
 \caption{Pitch Histogram for Example 1}
 \label{fig:example}
\end{figure}

We smooth the histogram in order to find the exact peak of each mountain by finding the moving average to smooth out short-term fluctuations and highlight the general trend of the data and we get a semi-Gaussian curve.  Choosing reasonable peaks is very crucial. To find the peak of each mountain, which represents the median of the performed pitch for each "note", we need to find the range of the mountain which is itself a challenging task and is discussed in [11].  After finding the range of each mountain, we model each mountain by a tilted Gaussian curve so that we can find a better peak. We fit the following curve to our data to find the parameters $c_1,\ldots,\ c_5$
\begin{equation}
{y=c}_1+c_2x+c_3e^{-(x-c_4)^2/c_5}
\end{equation}

 After finding the peak for each mountain in the histogram of audio, we have the frequency of every note in each sentence in cents. Since we only want reliable estimates, we have picked the mountains that have at least certain heights, so that we have enough data for fitting a Gaussian curve. Figure 3 shows the result of our Gaussian peak fit for the first significant peak of the pitch histogram in Example 1.
 
 \begin{figure}
 \centerline{
 \includegraphics[width=0.9\columnwidth]{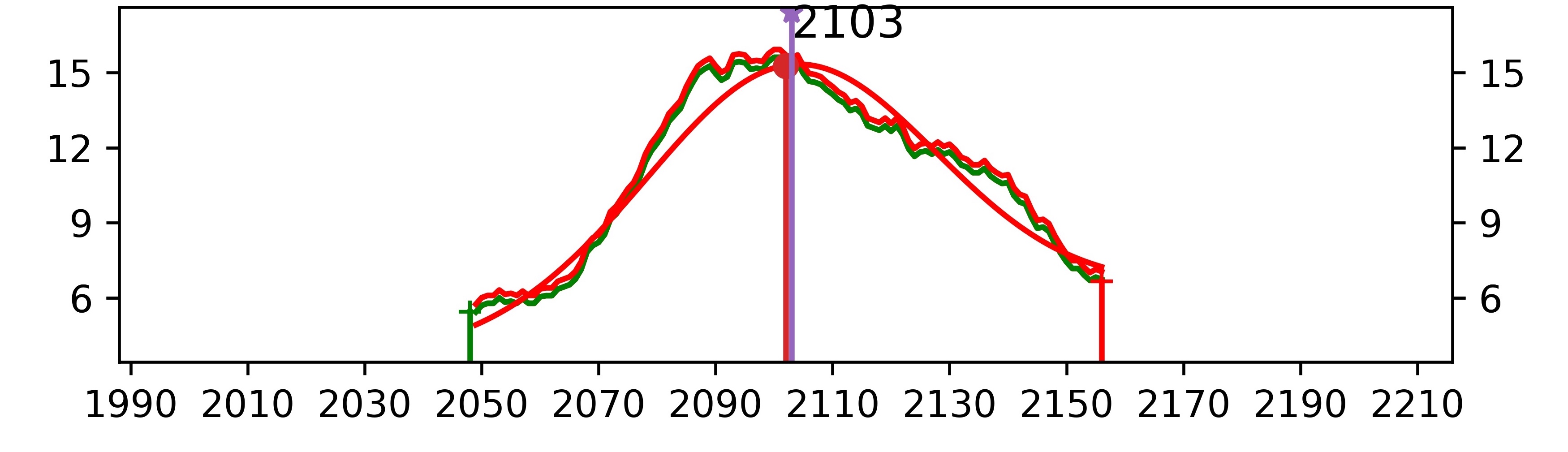}}
 \caption{Gaussian peak fit for the first significant peak of the pitch histogram Example 1}
 \label{fig:example}
\end{figure}

\section {Clustering the Main Peaks using DBSACN}

To see the pattern of drift in various notes throughout the piece, we first sort the frequency of the notes for each sentence. Assume that we have $n$ sentences $\{s_1,...,s_n\}$. For each sentence $s_i, 1<i<n$, we will have a number of frequency peaks $i_k$: $p_{i_1},...,p_{i_k}$. Then we need to cluster all these frequency points $p_{i_k}$ for $\forall {i,k}$, so that we have all the  fundamental frequencies associated to a pitch throughout the performance. Figure 4 shows the main detected frequencies for the sentences 0-15 in Example 1. The horizontal axis shows the sentence number and the vertical axis shows pitch (in cents).

 \begin{figure}
 \centerline{
 \includegraphics[width=0.9\columnwidth]{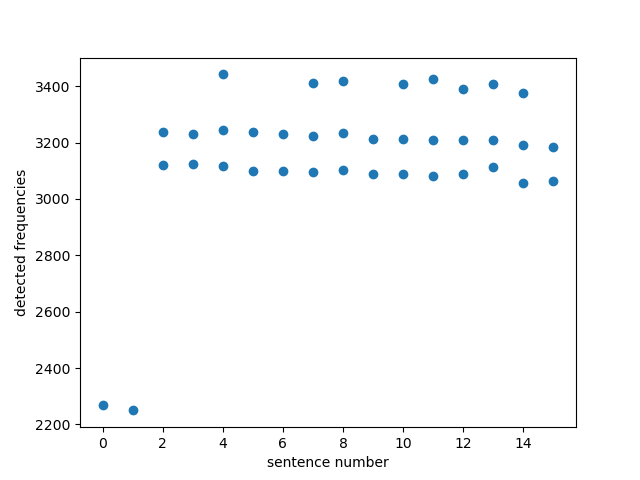}}
 \caption{The main detected frequencies for each sentence in Example 1}
 \label{fig:example}
\end{figure}

In order to cluster the frequency data points we used Density-based spatial clustering (DBSCAN) from scikit in Python. We then use linear regression to model each cluster. Figure 5 shows the result of our code for Example 1.  As can be seen in the figure, we have detected 4 cluster of points, each corresponds to a note. Each cluster is marked in the graph with a separate color. The first cluster (dark red color) only has two points, which is not significant. The other three clusters We then use linear regression to see the changes inside each cluster of frequency points. The slope of the lines in the linear regression shows the pattern of change in the tuning during the course of the performance.

\begin{figure}
 \centerline{
 \includegraphics[width=0.9\columnwidth]{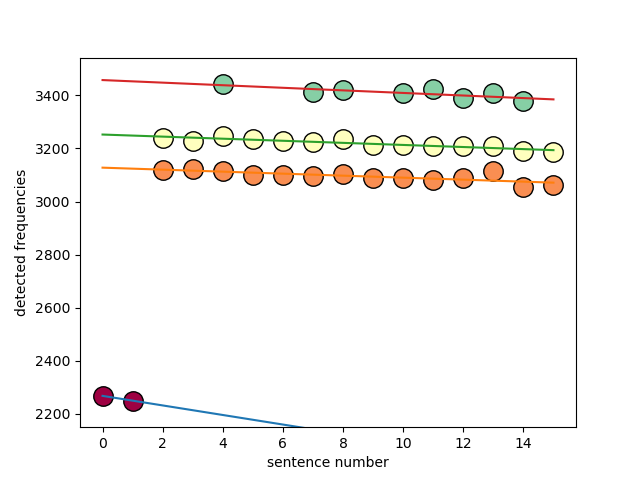}}
 \caption{Clustering of the frequencies using DBSCAN, and modeling each cluster with linear regression}
 \label{fig:example}
\end{figure}

\section {Acknowledgements}

Some parts of this paper is based on the author’s dissertation under the supervision of Prof. Blum at the Graduate Center, City University of New York [6].

\bibliography{ISMIRtemplate}

\end{document}